\documentclass[sigconf]{acmart}
\usepackage[linesnumbered, ruled, vlined]{algorithm2e}
\usepackage{algpseudocode}
\usepackage{amsmath}

\usepackage{amssymb}
\usepackage{graphicx}
\usepackage{subcaption}
\usepackage{tabularx}
\usepackage{multirow}
\usepackage{diagbox}
\usepackage{natbib}
\usepackage{bm} 

\AtBeginDocument{%
  }

\copyrightyear{2026}
\acmYear{2026}
\setcopyright{cc}
\setcctype{by}
\acmConference[AST '26]{7th ACM/IEEE International Conference on Automation of Software Test (AST 2026)}{April 13--14, 2026}{Rio de Janeiro, Brazil}
\acmBooktitle{7th ACM/IEEE International Conference on Automation of Software Test (AST 2026) (AST '26), April 13--14, 2026, Rio de Janeiro, Brazil}
\acmDOI{10.1145/3793654.3793751}
\acmISBN{979-8-4007-2476-3/2026/04}
\begin{document}


\title[A Hybrid Heuristic-Guided Deep Representation Arch. for Predicting Latent Zero-Day Vulnerabilities in Patched Functions]
{HYDRA: A Hybrid Heuristic-Guided Deep Representation Architecture for Predicting Latent Zero-Day Vulnerabilities in Patched Functions}


\author{Mohammad Farhad}
\affiliation{%
  \institution{School of Computing \& Informatics \\University of Louisiana at Lafayette}
  \city{Lafayette}
  \state{Louisiana}
  \country{USA}
}
\email{mohammad.farhad1@louisiana.edu}

\author{Sabbir Rahman}
\affiliation{%
  \institution{School of Computing \& Informatics \\University of Louisiana at Lafayette}
  \city{Lafayette}
  \state{Louisiana}
  \country{USA}
}
\email{sabbir.rahman1@louisiana.edu}
\author{Shuvalaxmi Dass}
\affiliation{%
  \institution{School of Computing \& Informatics \\University of Louisiana at Lafayette}
  \city{Lafayette}
  \state{Louisiana}
  \country{USA}
}
\email{shuvalaxmi.dass@louisiana.edu}
\renewcommand{\shortauthors}{Mohammad, et al.}

\begin{abstract}
Software security testing, particularly when enhanced with deep learning models, has become a powerful approach for improving software quality, enabling faster detection of known flaws in source code. However, many approaches miss post-fix latent vulnerabilities that remain even after patches typically due to incomplete fixes or overlooked issues may later lead to zero-day exploits. In this paper, we propose \textbf{HYDRA}, a \textit{\textbf{Hy}brid heuristic-guided \textbf{D}eep \textbf{R}epresentation \textbf{A}rchitecture} for predicting latent zero-day vulnerabilities in patched functions that combines rule-based heuristics with deep representation learning to detect latent risky code patterns that may persist after patches. It integrates static vulnerability rules, GraphCodeBERT embeddings, and a Variational Autoencoder (VAE) to uncover anomalies often missed by symbolic or neural models alone. We evaluate HYDRA in an unsupervised setting on patched functions from three diverse real-world software projects: Chrome, Android, and ImageMagick. Our results show HYDRA \textbf{predicts} 13.7\%, 20.6\%, and 24\% of functions from Chrome, Android, and ImageMagick respectively as containing latent risks, including both heuristic matches and cases without heuristic matches (\texttt{None}) that may lead to zero-day vulnerabilities. It outperforms baseline models that rely solely on regex-derived features or their combination with embeddings, uncovering truly risky code variants that largely align with known heuristic patterns. These results demonstrate HYDRA’s capability to surface hidden, previously undetected risks, advancing software security validation and supporting proactive zero-day vulnerabilities discovery. 
\end{abstract}


  



\maketitle
\section{INTRODUCTION}
Software vulnerabilities have emerged as a constant and dynamic threat as a result of the complexity and inter-connectivity of modern software system \cite{makrakis2021, ZIO2016137}. In response, software security testing has long been a cornerstone of software quality assurance, providing systematic and repeatable methods to verify software correctness, robustness, and reliability. As software systems continue to evolve rapidly, security testing has become essential for ensuring resilience and reliability throughout the software development lifecycle \cite{kamal2020risk, tuteja2012research, 8389562}. Despite these advancements in security testing and validation, zero-day vulnerabilities continue to pose a biggest risks to software security that are unpatched at the time of discovery and unknown to vendor. More formally, \textbf{Zero-day vulnerabilities} refer to previously unknown security flaws in software that are exploited before developers have had an opportunity to identify, patch, or disclose them \cite{bilge2012before}. These vulnerabilities often arise due to overlooked edge cases, incomplete patches, or misunderstandings of complex system behavior. As no official fix exists at the time of exploitation, zero-day attacks pose significant risks particularly in critical infrastructure, IoT, and open-source ecosystems. High profile incidents such as Stuxnet in  2010 \cite{stuxnet} and SolarWinds Orion in 2020 \cite{solarwinds} underscore the destructive potential of zero-day exploits and the importance of proactive vulnerability auditing. They enable hackers to take advantage of vulnerabilities before remedy is available, giving rise to silent, targeted attacks with devastating effects on users and systems \cite{Roumani, Zakharov}. Open-source companies sometimes decide to patch vulnerabilities covertly without revealing them through formal advisories like CVE \cite{cve} reports because of worries about maintainability and reputation. By converting zero-day risks into so called n-day vulnerabilities, this silent patching technique allows adversaries to target unpatched installations and reverse-engineer patches before downstream consumers can apply updates \cite{wang2019}. However most traditional methods like static analysis tools (e.g., Fortify) and signature based scanners (e.g. CodeQL) are built to identify known patterns of insecure code, despite the fact that researchers have made impressive strides in creating deep learning based vulnerability detectors.

\indent Using a variety of techniques, researchers have tried to find these hidden defects. For instance anomaly detection and autoencoder based techniques have been investigated to capture latent behavioral patterns of zero-day threats in network traffic \cite{nandakumarzero}, while machine learning classifiers have been trained to differentiate between benign code changes and secret security patches \cite{wang2019}. To describe patch timescales and programming language risks, others have been conducted empirical analysis of open source software repositories \cite{Zakharov}. However, the majority of these methods either use post-disclosure datasets or target network traffic or binary code, losing the proactive chance to reason about already fixed functions. In order to identify potential residual vulnerabilities, recent studies have emphasized the importance and practicality of mining silent patches (i.e., fixes that are applied without public disclosure) and tracing them back to their root causes or original vulnerable code.

\indent For instance, GraphSPD \cite{GraphSPD} and GRAPE \cite{Grape} use code property graphs and Graph Convulational Networks to detect undocumented vulnerability patches by analyzing structural differences. While effective, they frequently make the assumption that both pre-patch and post-patch code are available.
DeepDFA \cite{DeepDFA} extends vulnerability semantics using control/data-flow graphs and deep learning, achieving strong generalization and fast training, but still demands substantial structural modeling. Generative models like VulRepair \cite{vulrepair} and Zero-Shot Repair \cite{pearce} utilize large pre-trained language models (e.g., T5, Codex) to rewrite buggy code into secure versions. However, these methods focus on repair rather than detection and often depend on test suites, human oversight, or complex prompts. In contrast, data centric approaches such as VULGEN \cite{vulgen} and VGX \cite{vgx} generate synthetic vulnerable code via contextual transformations, while UL-VAE \cite{ul-vae} applies unsupervised anomaly detection to identify zero-day patterns in IoT malware. CLNX \cite{clnx} improves CodeBERT's robustness by incorporating commit context, and augmentation strategies help reduce overfitting in models like GraphCodeBERT.

\indent Despite these advances, most existing methods operate on vulnerable code snapshots, commit diffs, or symbolic traces of unpatched code. In contrast, our work explores whether patched functions assumed secure may still contain latent indicators of vulnerabilities, posing risks of re-emergent or zero-day exploits. In this work, we pose a central question: \textit{Can latent vulnerability patterns persist in patched functions previously considered secure thereby introducing future zero-day risks?} To investigate this, we introduce \textbf{HYDRA}, a hybrid prediction framework to improve post-patch software security testing, that combines domain-driven heuristic rules with semantic embeddings from a pre-trained model.

Motivated by prior findings on hidden or misclassified patches in open-source software \cite{wang2019} and the significance of silent fixes in long term code maintenance \cite{Zakharov}, we hypothesize that many patches may leave behind residual vulnerabilities either due to incomplete remediation or rushed developer fixes. HYDRA is designed to uncover these overlooked risks by integrating two complementary components: \textbf{(a). Symbolic pattern matching} based on five manually defined heuristic rules implemented using regular expressions (regex), and \textbf{(b). Semantic embeddings} generated from GraphCodeBERT \cite{graphcodebert}, which captures both token level context and data flow graph (DFG) information.
While regex offers interpretability and efficiency long used in static analysis to flag insecure code, as seen in VGX \cite{vgx} it is inherently limited to predefined patterns. In contrast, GraphCodeBERT enables HYDRA to generalize beyond symbolic rules, learning deeper semantic cues that may indicate risky behavior even when explicit patterns are absent. Notably, our model can label functions as \texttt{None} when no heuristic rule matches, highlighting previously patched function segments that may contain unknown or emergent zero-day risks. By combining symbolic reasoning with deep contextual understanding, HYDRA achieves both transparency in prediction and the generalization capacity of modern language models offering a novel perspective on post-patch vulnerability auditing.

In summary, this paper makes the following \textbf{ key contributions}:

\begin{itemize}
    \item We highlight the challenge of predicting latent zero-day vulnerabilities through common heuristic patterns that persist in previously patched functions a critical blind spot often overlooked by both static analysis and deep learning approaches. (\textbf{Section} \ref{sec:back})
    \item We propose HYDRA, a novel hybrid unsupervised architecture that integrates symbolic heuristic rules with deep GraphCodeBERT embeddings and a Variational Autoencoder (VAE), enabling effective prediction and clustering of residual risky code patterns without reliance on external program artifacts like ASTs or diff logs. (\textbf{Section} \ref{sec:method})
    \item We evaluate HYDRA on patched functions from Chrome, Android, and ImageMagick, showing its ability to detect latent vulnerabilities achieving up to 4× higher Silhouette, 10.1× higher CHI, and 72.2\% lower DBI than baselines while flagging 13–24\% of functions as potentially vulnerable, including those without explicit rule matches. (\textbf{Section} \ref{sec:eval})
    \item We further demonstrate HYDRA’s practical utility by identifying risky patterns through two real-world case studies from the Chrome and ImageMagick projects. (\textbf{Section} \ref{sec:dis})
\end{itemize}
Sections \ref{sec:val}, \ref{related_work} and \ref{sec:conc} mention about the threats to validity, related work and conclusion along with future work respectively. 
\section{BACKGROUND \& MOTIVATION}
\label{sec:back}
This section presents background and motivating examples that highlight the limitations of traditional tools such as static analyzers and CVE trained classifiers (see related work \ref{related_work}) in predicting residual vulnerabilities in source code. These examples represent the risky coding patterns that expose gaps in identifying post-patch risks, which HYDRA later revisits to demonstrate its ability to uncover, paving the way for zero-day vulnerability prediction.
\vspace{-0.2cm}
\subsection{Background}
Since downstream users might not be aware of unresolved or partially repaired vulnerabilities, latent vulnerability patches those done without disclosure, pose serious risks \cite{sun}. This motivates post-fix analysis frameworks that can verify the integrity of already fixed patches. Reducing the attack surface for unknown or new threats requires identifying code that has been patched but still has latent weaknesses. Even though structural embeddings can help identify unusual behaviour \cite{vda}, however, there is currently no method, to the best of our knowledge, for identifying latent risk vectors in patched functions that combines learned semantic embeddings with domain heuristics.\\ 
\indent In manual code review and static analysis, rule based heuristics such as race conditions, unsafe memory allocation, and missing null checks etc have long been used to find recurring patterns in a code \cite{Cppcheck, Flawfinder}. 
Each of these heuristics is associated to well known Common Weakness Enumerations (CWEs) IDs \cite{Cwe}, a list of common software and hardware weaknesses that can lead to vulnerabilities such as CWE-476 (NULL Pointer Deference) or CWE-119 (Buffer Overflow). These heuristics also frequently appear in Common Vulnerabilities and Exposures (CVEs) \cite{cve}, a system that provides a standardized identifier for publicly known vulnerabilities, facilitating traceability and comparison across projects indicating their critical relevance in real-world security patches. In order to improve code semantic representation, GraphCodeBERT \cite{graphcodebert} is a cutting edge pretrained transformer model that combines source code with data flow graph (DFG) structures. It allows structural reasoning without explicit CFG, AST inputs by internally constructing and encoding DFGs, even when we simply input source code. HYDRA introduces hybrid design that merges handcrafted vulnerability patterns with the representational strength of pre-trained models like GraphCodeBERT.\\
\vspace{-0.6cm}
\subsection{Motivating Examples and Challenges}
\begin{figure*}[ht]
  \centering
  \setlength{\fboxsep}{2pt} 
  \setlength{\fboxrule}{0.5pt} 

  \begin{subfigure}[b]{0.25\textwidth}
    \fcolorbox{gray!60}{white}{
      \includegraphics[width=6.3cm, height=2.8cm]{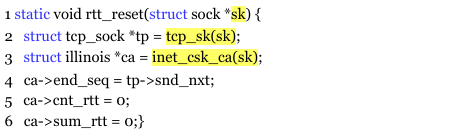}
    }
    \caption{Missing Null Check}
    \label{fig:missing_null_check}
  \end{subfigure}
  \begin{subfigure}[b]{0.3\textwidth}
    \fcolorbox{gray!60}{white}{
      \includegraphics[width=6.3cm, height=2.8cm]{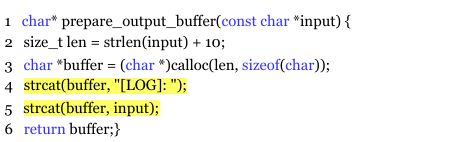}
    }
    \caption{Unsafe Memory Allocation}
    \label{fig:unsafe_memory_allocation}
  \end{subfigure}
  \begin{subfigure}[b]{0.3\textwidth}
    \fcolorbox{gray!60}{white}{
      \includegraphics[width=5.8cm, height=2.8cm]{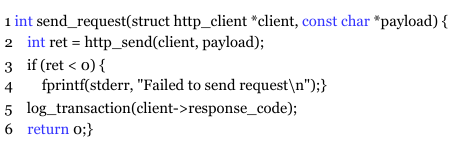}
    }
    \caption{Logging Without Halting}
    \label{fig:logging_without_halting}
  \end{subfigure}

  \vspace{0.6cm}

  \begin{subfigure}[b]{0.3\textwidth}
    \fcolorbox{gray!60}{white}{
      \includegraphics[width=5.5cm, height=2.8cm]{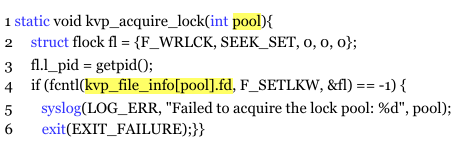}
    }
    \caption{Missing Bounds Check}
    \label{fig:missing_bounds_check}
  \end{subfigure}
  \hspace{0.3cm}
  \begin{subfigure}[b]{0.3\textwidth}
    \fcolorbox{gray!60}{white}{
      \includegraphics[width=5.5cm, height=2.8cm]{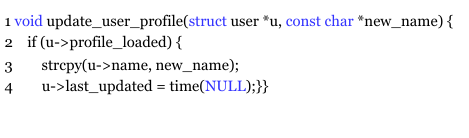}
    }
    \caption{Race Condition}
    \label{fig:race_condition}
  \end{subfigure}
    \vspace{-0.3cm}
  \caption{Illustrative Examples of Common Risky Heuristic Patterns Utilized in HYDRA.}
  \Description{Illustrative examples of heuristic-based risky code patterns used by HYDRA. The subfigures depict common post-fix vulnerabilities, including missing null checks, unsafe memory allocation, logging without halting execution, missing bounds checks, and race conditions that may persist in patched functions.}

  \label{fig:heuristic_examples}
  \vspace{-0.4cm}
\end{figure*}

HYDRA is motivated by the frequent occurrence of residual risky heuristic patterns in post-patch functions. We present \textbf{five} main heuristic rules derived from most common insecure coding patterns observed in real-world OSS and CVE linked patches \cite{nvd}. While HYDRA currently employs these five heuristics, its module is designed to be extensible, allowing incorporation of additional rules for broader generalization. In our work, we chose these five motivating examples as these rules capture flaws most often introduced by human error, poor fix localization, or misinterpretation of vulnerability contexts \cite{linux-kernel}, hence they reveal key challenges associated with residual risky patterns in patched functions. They motivate our hybrid approach by revealing persistent risk patterns even in patched function. Though developed independently from empirical observations, these rules later aligned with major Linux Kernel vulnerability classes \cite{linux-kernel}, such as buffer overflows and null pointer dereferences underscoring their practical relevance and generality. The risky heuristic patterns and associated challenges are defined below:\\
\textbf{Missing Null Check:} In Figure \ref{fig:missing_null_check}, the \texttt{rtt\_reset} function in line 1 accepts a pointer $sk$ and immediately passes it into transformations functions \texttt{tcp\_sk(sk)} and \texttt{inet\_csk\_ca(sk)} (see line 2 and 3). The result of those functions (tp, ca) are used without any nullity verification. This dereferenced pointers without null checks, which is a common bug type in real-world vulnerability datasets (e,g, CWE-476). This leads to \textit{information disclosure} and \textit{denial of service (DoS)}, where dereferencing unchecked pointers can allow unintended access to sensitive memory regions or privileged operations.


\noindent \textbf{Unsafe Memory Allocation:} The use of \texttt{calloc} (line 3) in Figure \ref{fig:unsafe_memory_allocation}, without checking its return value introduces risk of dereferencing a null pointer if memory allocation fails. If \texttt{calloc} returns \texttt{NULL} and the subsequent \texttt{strcat} (line 4 and 5 highlighted) operation is invoked on this null pointer, it will lead to undefined behavior, likely causing segmentation faults or process crashes. Such patterns align with \textit{CWE-690: Unchecked Return Value to NULL Pointer Dereference}, which emphasizes the importance of validating the success of memory allocation before using the pointer. It leads to \textit{memory corruption and manipulations}, where unchecked allocation results can cause undefined behavior, data overwrites or exploitation through crafted inputs.

\noindent \textbf{Logging Without Halting:} The function \texttt{fprintf(stderr, ...)} (line 4) attempts to log an error condition when a configuration file fails to open in Figure \ref{fig:logging_without_halting}. However, it does not perform any blocking or termination behavior afterward such as returning an error code or exiting the program thus allowing execution to proceed in an invalid or undefined state. This may result in dereferencing \texttt{client->response\_code} which may be uninitialized or invalid, depending on the \texttt{http\_send} outcome. This issue corresponds to \textit{CWE-703: Improper Check or Handling of Exceptional Conditions}, where an error is acknowledged but not adequately acted upon, leading to continued use of an invalid program state. Due to this, might happen \textit{privilege escalation}, as the program continues execution in invalid states after logging errors, potentially leaking sensitive data or triggering crashes.

\noindent \textbf{Missing Bounds Check:} In Figure \ref{fig:missing_bounds_check}, line 4 (highlighted) accesses \texttt{kvp\_file\_info[pool]}, referencing an array or struct pointer. However, the function lacks validation to ensure that pool is within bounds (e.g.,\texttt{0 $\le$ pool < MAX\_POOLS}) or that \texttt{kvp\_file\_info[pool].\newline fd} is a valid file descriptor (e.g.,$\geq 0$). If pool is not properly validated before invocation, it may lead to undefined behavior or memory corruption. Moreover, using an invalid \texttt{fd} in \texttt{fcntl} (line 5) could result in runtime errors, particularly if the value is uninitialized posing a risk consistent with CWE-125 (Out-of-Bounds Read). This conducts to \textit{buffer overflows} and \textit{injection attacks}, enabling attackers to overwrite memory or inject malicious payloads via invalidated array or pointer access.

\noindent \textbf{Race Condition:} In Figure \ref{fig:race_condition}, the \texttt{update\_user\_profile} function (line 1) modifies a shared user object without any synchronization. In multi-threaded environments, concurrent invocations on the same user instance can lead to unsafe interleaving e.g., during the \texttt{strcpy} operation (line 3) or while updating \texttt{last\_updated}. This may result in inconsistent logs, corrupted state, or memory safety issues depending on subsequent use. The lack of synchronization primitives (e.g., \texttt{pthread\_mutex\_lock}) constitutes a classic data race, aligning with CWE-362: Concurrent Execution using Shared Resource with Improper Synchronization. It may operates to \textit{deadlocks} and \textit{inconsistent system states}, as lack of synchronization in multi-threaded environments may corrupt shared data or halt program progress.
\vspace{-0.3cm}
\section{METHODOLOGY}
\label{sec:method}
This section outlines the proposed approach of HYDRA. After giving a general introduction to HYDRA, we go into details of each module. 
\vspace{-0.4cm}
\subsection{Overview of HYDRA} 
Identifying latent vulnerabilities especially those that persist after patching is a critical yet underexplored challenge in software security \cite{Grape, GraphSPD, Roumani, sun}. While many detection systems leverage deep neural models to learn from source code \cite{vda, svulD, GraphSPD, Devign}, few tackle the nuanced task of predicting future zero-day risks in patched code, where rushed fixes or incomplete understanding may leave exploitable remnants.
To address this, we propose \textbf{HYDRA} (\textit{\textbf{Hy}brid heuristic-guided \textbf{D}eep \textbf{R}epresentation \textbf{A}rchitecture}), a novel hybrid framework that bridges deep semantic reasoning with symbolic pattern explainability. HYDRA integrates \textit{regex based heuristic prediction} with \textit{GraphCodeBERT} \cite{graphcodebert} to capture both structural and contextual code semantics. An unsupervised clustering module (e.g., K-Means, VAE) further enhances prediction by surfacing outlier patterns potential indicators of unknown or non obvious risks beyond predefined rules. We will now describe the HYDRA architetcure in detail. 

\vspace{-0.4cm}
\subsection{The HYDRA Architecture} 
\begin{figure*}[htbp]
  \centering
  \includegraphics[width=17cm, height=5cm]{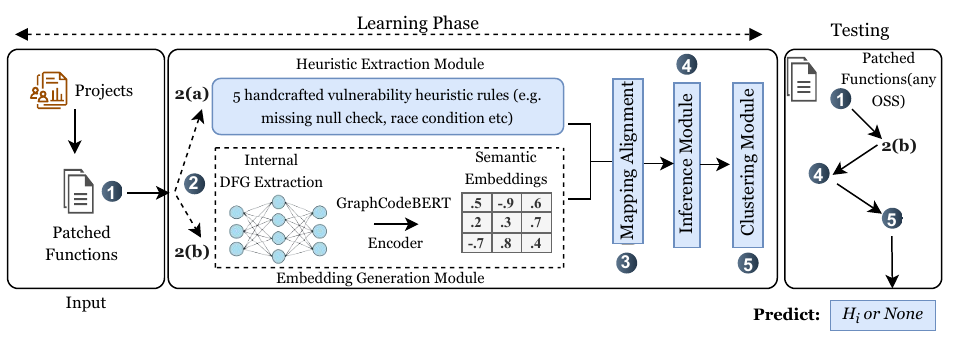}
  \vspace{-0.8cm}
  \caption{The proposed architecture of HYDRA.}
  \Description{The diagram shows the HYDRA architecture for predicting latent zero-day vulnerabilities in patched functions. The system combines static heuristic-based vulnerability rules with GraphCodeBERT representations and a variational autoencoder to identify latent risky code patterns that may persist after patches.}
  \label{fig:HYDRA-Architecture}
\end{figure*}

HYDRA is a hybrid vulnerability prediction framework designed to identify latent zero-day risk indicators in already patched functions. As shown in Figure \ref{fig:HYDRA-Architecture}, it has two phases: Learning and Testing. \newline
\noindent \textbf{I. Learning Phase.} In this phase, HYDRA ingests post-fix functions which are passed through two parallel process pipelines: one \textit{heuristic-driven} and the other based on \textit{semantic embeddings}. The Learning pipeline consists of the following steps:
\vspace{-0.01in}

\begin{enumerate}
    \item \textbf{Input:} A set of fixed (patched) functions written in C, collected from a publicly available vulnerability patch dataset, BigVul.
    \item \textbf{Heuristic Feature Extraction Module:} Patched functions from input is parsed here using pattern-matching rules (e.g. regex) to detect the presence of known insecure coding practices (e.g. race condition, unsafe memory allocation) that leads to future zero-day attacks.
    \item \textbf{Semantic Embedding Module:} Simultaneously, the input patched functions are also passed  to the pre-trained GraphCodeBERT for tokenization, which produces a high dimensional representation of the function's semantics based on both code tokens and implicit data-flow-graphs (DFG).
    \item \textbf{Mapping Alignment:} The corresponding heuristic vector of each functions is paired with its GraphCodeBERT embedding to form  a training tuple (Embedding $\,\to\,$ heuristic vector).
    \item \textbf{Inference Module:} This module fuses the 768-dimensional GraphCodeBERT embedding with a 5 heuristic vector during training, forming a 773-dimensional input that is passed through a Variational Autoencoder (VAE) to learn compact latent representations. During testing, only the 768-dimensional embeddings are used to project unseen functions into the same latent space.
    \item \textbf{Clustering Module:} This module applies K-Means clustering on the latent representations learned by the VAE to group functions into vulnerability clusters and assign labels based on the proximity to heuristic traits or \texttt{None} regions.
\end{enumerate}

Let's dig deeper into (2), (3), (4) and (5) modules.
\vspace{-0.2em}
\subsubsection{Heuristic Feature Encoding}
To capture interpretable, domain specific vulnerability signals, HYDRA incorporates a handcrafted heuristic extraction module implemented in python. This component is grounded in empirical security research and established CWE categorizations, particularly targeting vulnerability root causes frequently encountered in post-patch software. We design the following high impact binary vulnerability signatures:
\begin{itemize}
    \item $\boldsymbol{H_1}$: Missing null pointer check ($\to$ may lead to CWE-476). 
    \item $\boldsymbol{H_2}$: Race condition check due to lack of synchronization ($\to$ CWE-362).  
    \item $\boldsymbol{H_3}$: Missing bounds check on buffer/data ($\to$ CWE-119, CWE- 120). 
    \item $\boldsymbol{H_4}$: Unsafe memory allocation check for $malloc/calloc/realloc$ failures ($\to$ CWE-690). 
    \item $\boldsymbol{H_5}$: Logging errors without control flow blocking check (e.g. missing return or exit) ($\to$ CWE-390, CWE-703).
    \item \texttt{\textbf{None}}: Represents functions with no heuristic match (benign  functions or possibility to have unknown vulnerability). Not provided in input step. 
\end{itemize}
\noindent For each patched function, we extract features using handcrafted regular expressions. where each function emits a heuristic match vector: \begin{center}  $V_h \in \{0,1\}^5$ \end{center} where each bit in the vector indicates the presence (1) or absence (0) of a specific vulnerability signature. For example $V_h = [1, 0, 0, 1, 0]$ indicates that the function both logs errors unsafely and lacks a null pointer check. This feature vector serves a dual purpose: \textbf{(1).} providing interpretable evidence that complements black-box deep learning predictions. \textbf{(2).} serving as an auxiliary signal to improve recall on semantically simple yet security critical cases often missed by deep models.

\subsubsection{Embedding Extraction with GraphCodeBERT}
While the heuristic extractor flags known vulnerability patterns via regex, it cannot capture latent or semantically obfuscated flaws. To address this, HYDRA integrates GraphCodeBERT \cite{graphcodebert}, a pre-trained transformer encoder tailored for source code. It generates deep contextual embeddings of patched functions, enabling the model to generalize beyond surface syntax (\textit{code syntax only }) and detect subtle indicators of risky logic or incomplete fixes. GraphCodeBERT leverages both token sequences and data flow graphs (DFGs) to learn a rich, compositional representation of program behavior.\\
\indent Let $f_i$ denote a preprocessed function in C, drawn from the patched function samples $F = \{f_1, f_2,....f_n\}$. Each function $f_i$ is tokenized into sequence of subword tokens (T):
\begin{center}
    $T_i = \{t_{i1}, t_{i2},...t_{imi} \}, \; t_{ij} \; \in \; \sum $
\end{center}

where $\sum$ is the vocabulary space, and $t_{ij}$ is the j-th token of the i-th function. Simultaneously, GraphCodeBERT constructs associated data flow graph:
\begin{center}
    $G_i = \{\nu_i, \varepsilon_i \}, \; \nu_i \;  \subseteq T_i$
\end{center}
where nodes $\nu_i$ represents tokens that read/write shared variables, and edges $\varepsilon_i \subseteq$ $\nu_i$ x $\nu_i$ model variable dependencies. GraphCodeBERT then applies a dual learning objective to jointly optimize: \newline \textbf{(1).} \textit{Masked Language Modeling (MLM):} Predicting masked tokens from surrounding context. and \textbf{(2).} \textit{Edge Prediction (EP):} Predicting existence of data flow edges between token pairs.\\
\indent Through these objectives, a function $f_i$ is encoded into a high-dimensional vector space:
\begin{center}
    $E_s^{(i)} = GraphCodeBERT(T_i, G_i) \, \in \, R^{d}, \; d=768 $
\end{center}

Here, $d$ is dimensional real-valued vector space and $R$ is the mathematical notation for a $d$-dimensional vector of real numbers. Each $E_s^{(i)}$ serves as a semantic embedding that captures structural, control-flow, and data-flow properties of function beyond lexical tokens.\\
We denote the complete embedding set across all samples as:
\begin{center}
    $\epsilon_{s} = \{E_s^{1}, E_s^{2},...E_s^{n} \} $
\end{center}
These embeddings are passed as real valued inputs into the downstream hybrid model alongside binary heuristic vectors (see Section \ref{subsubsection:inference}). Empirically, we observe that GraphCodeBERT embeddings tend to organize functions into distinct semantic regions, where structurally or logically risky code patterns become separable in the high-dimensional embedding space.\\
\indent In HYDRA, GraphCodeBERT is used as a frozen encoder, no parameter updates are performed ensuring reproducibility and leveraging knowledge distilled from CodeSearchNet and CodeXGLUE during pretraining.

\subsubsection{Mapping Alignment and Inference Module}
\label{subsubsection:inference}

To unify symbolic vulnerability indicators with deep semantic representations, HYDRA employs a hybrid learning mechanism based on \textit{early fusion}, where heuristically derived features and pretrained code embeddings are concatenated into a single latent feature space. This formulation supports both explainability and generalization during vulnerability inference.

Let us define: 
\begin{center}
$V_h(x) = [v_1, v_2, \dots, v_5]$, where $v_i \in {0,1}$
\end{center}

\begin{itemize}
\item $x \in C$ denotes a patched C function.
\item $V_h(x) \in {0,1}^5$ is a binary vector indicating the presence of each heuristic pattern (e.g., $v_1 = 1$ if a null check is missing).
\item $E_s(x) \in \mathbb{R}^{768}$ is the semantic embedding of $x$ obtained using GraphCodeBERT.
\end{itemize}

The fused representation is defined as:
\begin{center}
$z(x) = \text{Concat}(V_h(x), E_s(x)) \in \mathbb{R}^{773}$
\end{center}

\noindent To learn more compact latent vulnerability structure, we use a Variational Autoencoder (VAE) to project $z(x)$ into a compact latent space. \newline
\textit{Clustering Module.} We then apply K-Means clustering over the learned representations to group functions with similar vulnerability characteristics and assign risk labels from the set ${\texttt{H}_1, \texttt{H}_2, \dots, \texttt{H}_5, \texttt{None}}$ based on cluster level heuristic prevalence.\newline
\noindent \textbf{ II. Testing Phase.} In test phase, for an unseen patched function $X_{test}$: \textbf{(1).} If $argmax \; f_0(L) = None$, HYDRA interprets the fixed function as either secure so far or as potentially containing an unknown (zero-day) vulnerability, (i.e., no known heuristic pattern is detected). \textbf{(2).} Otherwise, if the output corresponds to one of the five defined heuristic rules, HYDRA identifies the patched function as still containing latent vulnerability indicators such as “missing null check” thus flagging it as a potential zero-day vulnerability with explainable context. In this phase, only the unseen patched source functions from OSS is passed through the DL model pipeline to predict the future latent vulnerabilities from the patched functions.\newline
\vspace{-0.2em}
\begin{algorithm}[t]
\small
\caption{HYDRA: Hybrid Deep Representation Architecture for Vulnerability Prediction on Patched Functions.}
\label{alg:hydra}
\KwIn{
    Patched function corpus $\mathcal{F} = \{f_1, f_2, \dots, f_n\}$\;
    Pre-trained GraphCodeBERT model $G$\;
    Heuristic rule set $\mathcal{H} = \{h_1, h_2, \dots, h_5\}$
}
\KwOut{
    Risk label vector $\mathcal{V} = \{v_1, v_2, \dots, v_n\}$, where $v_i \in \mathcal{H} \cup \{\texttt{None}\}$
}
\ForEach{$f_i \in \mathcal{F}$}{
    Apply all $h_j \in \mathcal{H}$ to $f_i$ via regex pattern matching\;
    Construct heuristic vector $H_i \in \{0,1\}^5$\;

    Encode $f_i$ using GraphCodeBERT:\;
    $E_i \leftarrow G(f_i)$, where $E_i \in \mathbb{R}^{768}$\;

    Concatenate: $Z_i = [E_i \, \| \, H_i] \in \mathbb{R}^{773}$\;

    Train a Variational Autoencoder (VAE) on $\{Z_1, Z_2, \dots, Z_n\}$ to learn compact and continuous latent representations $\{L_1, L_2, \dots, L_n\}$\;

    Cluster $\{L_i\}$ using K-Means to assign cluster labels $\{v_1, v_2, \dots, v_n\}$, where $v_i \in \mathcal{H} \cup \{\texttt{None}\}$\;
}
\Return{$\mathcal{V}$}
\end{algorithm}
\noindent The overall HYDRA model prediction is described in \textbf{Algorithm} \ref{alg:hydra}. Here each function $f_i$ is first analyzed using five heuristic rules (lines 2–3). GraphCodeBERT then encodes the function into a 768-dimensional embedding $E_i$ (lines 4–5), which is concatenated with the 5-dimensional heuristic vector $H_i$ to form a fused 773-dimensional vector $Z_i$ (line 6). This vector is passed through a Variational Autoencoder (VAE) to generate latent representations $L_i$ (line 7). Finally, K-Means clustering groups similar functions, and heuristic-based labels including \texttt{None} are assigned based on latent structure (lines 8–9).

\vspace{-0.2cm}
\section{EVALUATION OF HYDRA}
\label{sec:eval}
We evaluate the  effectiveness of HYDRA based on following research questions:
\begin{itemize}
    \item \textbf{RQ1:} Can HYDRA predict latent risky patterns (heuristic rules) in patched functions missed by symbolic or deep models that may lead to zero-day vulnerabilities?
    \item \textbf{RQ2:} How do HYDRA’s components influence the quality of learned representations for clustering patched functions?
    \item \textbf{RQ3:} Can HYDRA anticipate novel or previously unseen vulnerability patterns not explicitly covered during training?
\end{itemize}
\vspace{-0.2cm}
\subsection{Experimental Setup}
\textbf{Implementation}.
The deep learning components (including the integration of GraphCodeBERT and classifier models) are implemented using the PyTorch (v2.7.1) \cite{pytorch} and the HuggingFace Transformers (v4.37) library \cite{transformers}. The heuristic rule engine for feature extraction is implemented with custom regex modules and Python’s built-in AST parser for syntactic verification. All experiments are conducted on a multi-core Linux server equipped with Intel Core i7-12700F CPU (3.0GHz × 8 cores), and NVIDIA RTX 4070 GPU (12GB VRAM), and 16GB of RAM, running Kali 2023.2 with Linux kernel version 6.1. 

\noindent \textbf{HYDRA and Baseline Models.} To the best of our knowledge, there are no existing baseline models in unsupervised settings that specifically designed for post-patch vulnerability prediction at the time of developing this work. Hence, to assess HYDRA’s effectiveness in this unsupervised setting for vulnerability prediction, we construct multiple baseline design variant models, detailed as follows:
\begin{enumerate}

\item [\textbf{[M1]}] \textbf{Regex + K-means}: A naive most basic baseline model using only handcrafted regex-based heuristic rules, where K-means is applied to cluster functions based solely on their binary rule presence.
\item [\textbf{[M2]}]\textbf{GraphCodeBERT + K-means}: We extract semantic embeddings of the functions using GraphCodeBERT and apply K-means to identify high density regions of potential vulnerability patterns.
\item [\textbf{[M3]}] \textbf{Regex + GraphCodeBERT + K-means}: Combines the regex-based heuristic rule vectors with GraphCodeBERT embeddings via early fusion. The concatenated representation is directly clustered using K-means to uncover latent structure informed by both symbolic and semantic signals.
 \item[\textbf{[HYDRA]}]\textbf{Regex + GraphCodeBERT + VAE + K-means}: Our proposed architecture uses a Variational Autoencoder (VAE) trained on early fused GraphCodeBERT embeddings and heuristic rule vectors. At inference time, only the embeddings are provided to the VAE to project functions into a latent space, where clustering is applied to reveal potential risk patterns.
\end{enumerate}

\noindent \textbf{Dataset Construction.} We construct our dataset using Big-Vul \cite{Bigvul}, a large scale vulnerability dataset based on the National Vulnerability Database (NVD) \cite{nvd}, which provides real-world vulnerable and fixed code pairs. Specifically, we extract 20,451 patched C functions from the Linux project for training our model, sourced from CVE linked repositories \cite{cve}. These functions are treated as the fixed versions of previously vulnerable code making them ideal for studying residual risks and latent vulnerabilities that may still lead to future zero-day exploits.
To evaluate generalization, we test our models on patched functions from three diverse codebases: Chrome (16,387 functions), Android (2,322 functions), and ImageMagick (1,703 functions). We use these projects due to their diversity in application domains, code complexity, and real-world vulnerability history. \newline
\noindent \textbf{Evaluation Metrics.}
\label{sec:metrics}
Given the unsupervised nature of HYDRA, to assess the effectiveness of it’s latent representations in forming semantically meaningful clusters, we evaluate clustering quality using three standard unsupervised metrics Silhouette Score \cite{sil}, Calinski-Harabasz Index (CHI) \cite{calinski}, and Davies-Bouldin Index (DBI) \cite{davies}. Each metric quantifies the balance between intra-cluster cohesion and inter-cluster separation. The ranges for these scores  are $[-1, 1]$, $[0, \infty)$ and $[0, \infty)$ respectively.
Higher values for Silhouette and CHI indicate better clustering, while lower DBI values indicate better cluster separation and compactness. 
\vspace{-0.3em}
\subsection{RQ1: Predicting Latent Risky Heuristic Patterns by HYDRA}
\textbf{Method.} To assess HYDRA’s effectiveness in identifying latent risks in patched functions particularly those missed by symbolic (regex) or semantic (GraphCodeBERT) methods, we evaluate heuristic prediction performance across three model variants: M1, M3, and HYDRA. The model M2 (GraphCodeBERT $+$ KMeans) is not included here, as it produces purely semantic representations without any connection to the heuristic labels. HYDRA is explicitly designed to align both symbolic and semantic cues with risk patterns. The models are trained to associate these representations with predefined heuristic labels $H_{1}$ to $H_{5}$. During testing, they predict the presence of these risk patterns in patched functions drawn from three real-world projects: Chrome, Android, and ImageMagick. Detailed results for Chrome are shown in Table \ref{tbl:heuristic_type}, with full cross-project summaries provided in Table \ref{tbl:projects_heuristic_percentage}.\\
\textbf{Results.} Table \ref{tbl:heuristic_type} summarizes the heuristic rule prediction match counts for the Chrome test dataset across three model variants - M1, M3, and HYDRA - illustrating how total predicted matches are computed for a single project. While HYDRA detects fewer total matches (788) than M1 (1126) and M3 (921), this reduction reflects its focus on learning more compact and generalizable representations. For example, HYDRA captures a comparable number of $H_2$ matches (292), yet avoids potential over-prediction in $H_1$ and $H_3$, which are more susceptible to superficial matches in simpler models. Although M1 yields the highest count, its reliance on regex alone increases the likelihood of redundant or noisy predictions. HYDRA, by integrating symbolic and semantic features through a VAE-based encoder, emphasizes representational fidelity, producing more selective yet interpretable clusters. These results suggest that HYDRA enables conservative but meaningful heuristic alignment, improving generalization and prediction of residual risks in patched function. Notably, $H_4$ and $H_5$ — unsafe memory allocation and unsafe logging were not observed in Chrome. This likely reflects their rarity in real-world patches, where developers prioritize immediately visible issues. Additionally, such patterns often require deeper semantic inference or runtime context, which may not be evident in static code alone.
\vspace{-0.2cm}

\begin{table}[ht]
\centering
\footnotesize
\caption{ Heuristic rule prediction counts by model variants (M1, M3, HYDRA) on the Chrome test project of patched functions.}
\vspace{-0.3cm}
\label{tbl:heuristic_type}
\begin{tabular}{c | c | c | c}
 \hline
 \textbf{Heuristic Rules} & M1 & M3 & HYDRA\\ 
 \hline
 $H_1$ & 645 & 588 & 460\\ 

 $H_2$ & 292 & 278 & 292\\
 
 $H_3$ & 189 & 55 & 36\\
 
 $H_4$ & 0 & 0 & 0\\
 
 $H_5$ & 0 & 0 & 0\\

 \hline
 Total Matched Heuristic & 1126 & 921 & 788\\
 \hline
\end{tabular}
\vspace{-0.3cm}
\end{table}

\noindent \textbf{Cross project analysis.}
Table \ref{tbl:projects_heuristic_percentage} 
presents matched heuristic rule counts and their corresponding percentages across Chrome, Android, and ImageMagick for various model variants. The hybrid HYDRA model consistently yields the lowest match rates 4.80\% in Chrome, 11.1\% in Android, and 16.03\% in ImageMagick indicating stronger filtering of false positives while still capturing meaningful risk patterns. In contrast, models like M1 and M3 report significantly higher match rates (e.g., M1 shows 6.87\% in Chrome and 22.19\% in ImageMagick), suggesting over-matching due to weaker generalization. HYDRA’s integration of symbolic rules with latent code representations improves precision and enables more refined post-patch vulnerability prediction across diverse codebases. Its performance on both large-scale projects (e.g., Chrome) and smaller ones (e.g., ImageMagick) highlights the robustness and adaptability of the hybrid approach.\newline
\vspace{-0.3cm}

\vspace{-0.2cm}
\begin{table}[ht]
\centering
\footnotesize
\caption{Heuristic rule prediction summary for Chrome, Android, and ImageMagick across model variants. Values indicate total matches and corresponding percentages.}
\vspace{-0.2cm}
\label{tbl:projects_heuristic_percentage}
\begin{tabular}{c|  c|  c|  c}
 \hline
 \textbf{Model} & Chrome & Android & ImageMagick \\ 
 \hline
 M1 & 1126 (6.87\%) & 411 (17.7\%) & 378 (22.19\%)\\
 
 M3 & 921 (5.62\%) & 304 (13.1\%) & 351 (20.61\%)\\
 \textbf{HYDRA} & \textbf{788 (4.80\%)} & \textbf{257 (11.1\%)} & \textbf{273 (16.03\%)}\\
 \hline
\end{tabular}
\vspace{-0.3cm}
\end{table}
\vspace{-0.5cm}

\begin{figure*}[ht]
  \centering
   \begin{subfigure}[t]{0.30\textwidth}
    \includegraphics[width=5.3cm, height=3.3cm]{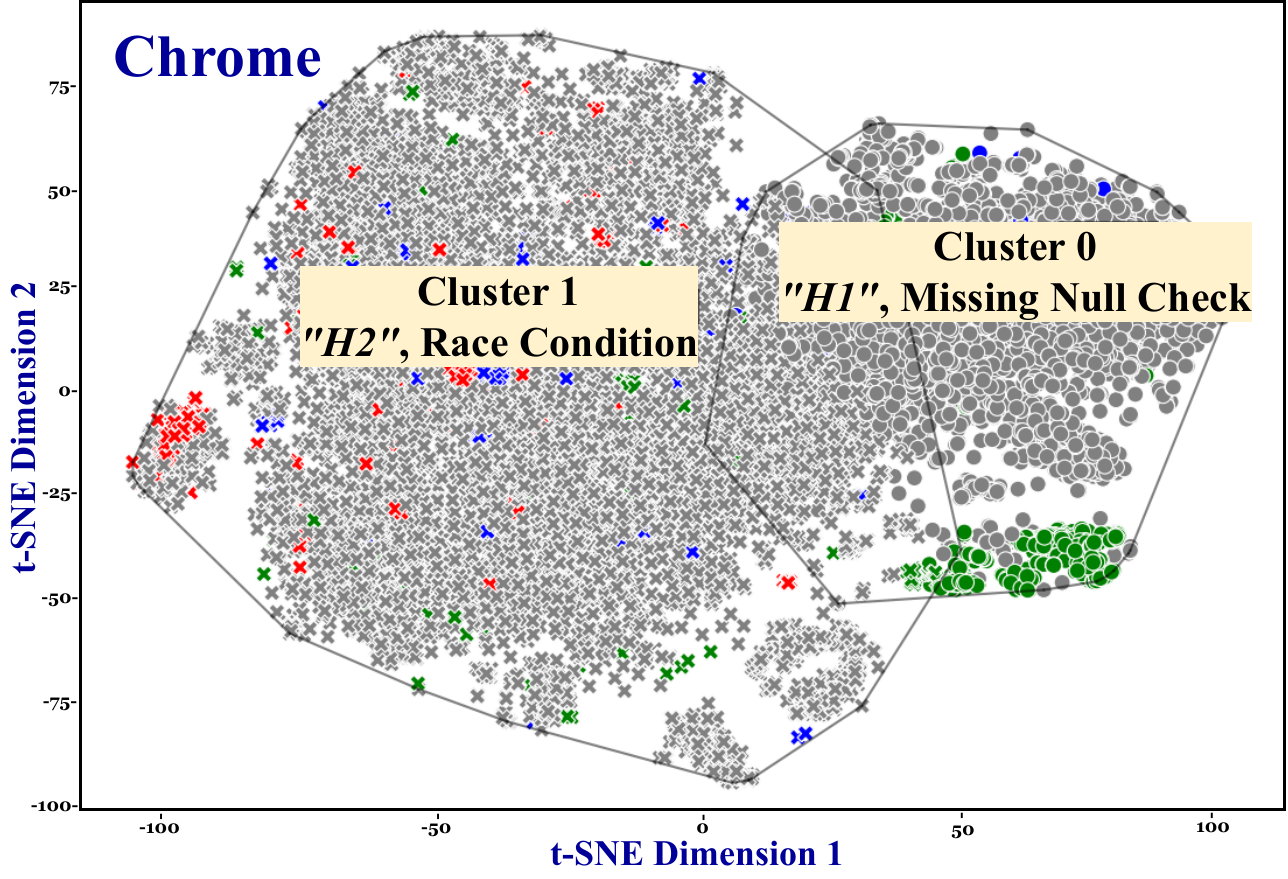}
    \label{fig:chrome_t-sne}
  \end{subfigure}
  \hspace{0.1cm}
  \begin{subfigure}[t]{0.30\textwidth}
    \includegraphics[width=5.3cm, height=3.3cm]{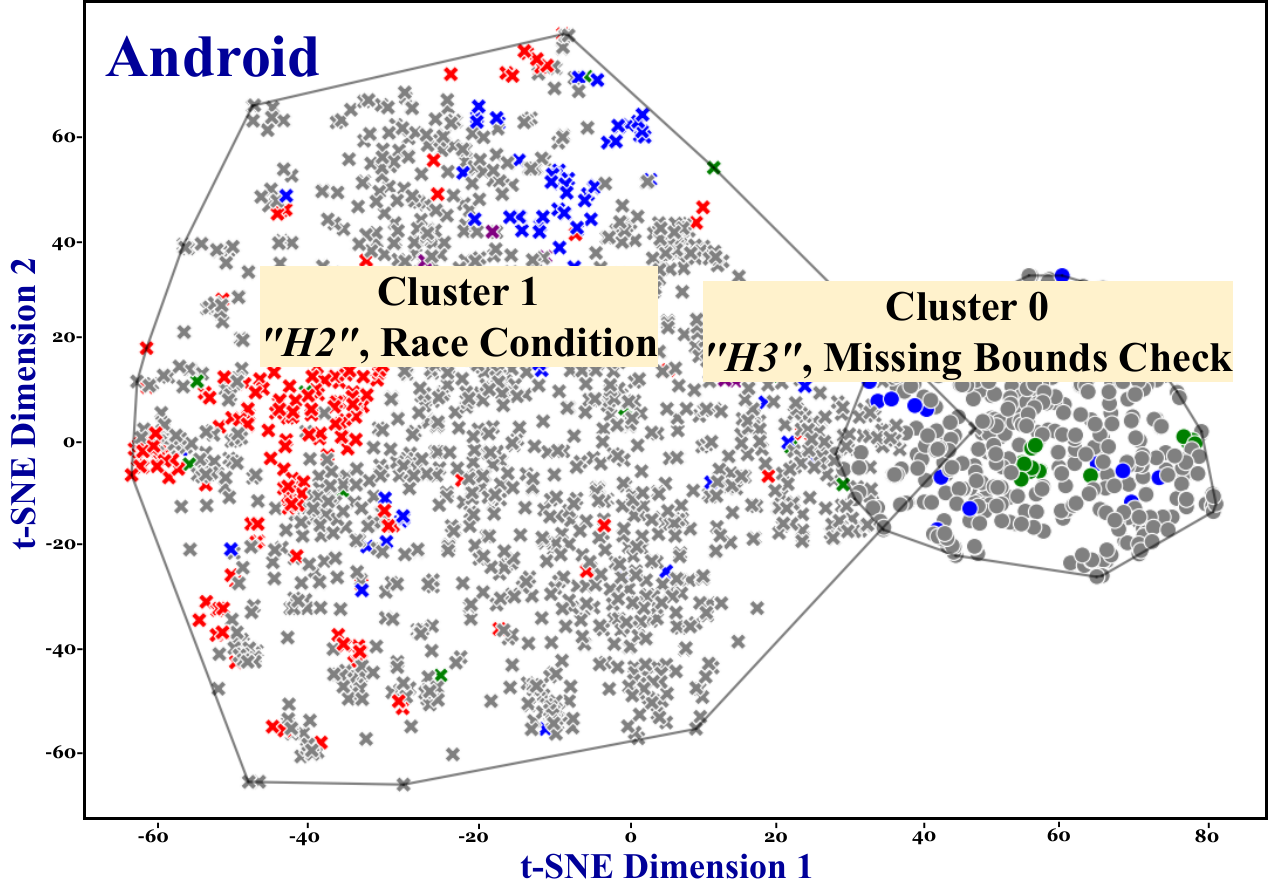}
    \label{fig:android_t-sne}
  \end{subfigure}
  \hspace{0.1cm}
  \begin{subfigure}[t]{0.30\textwidth}
    \includegraphics[width=5.3cm, height=3.3cm]{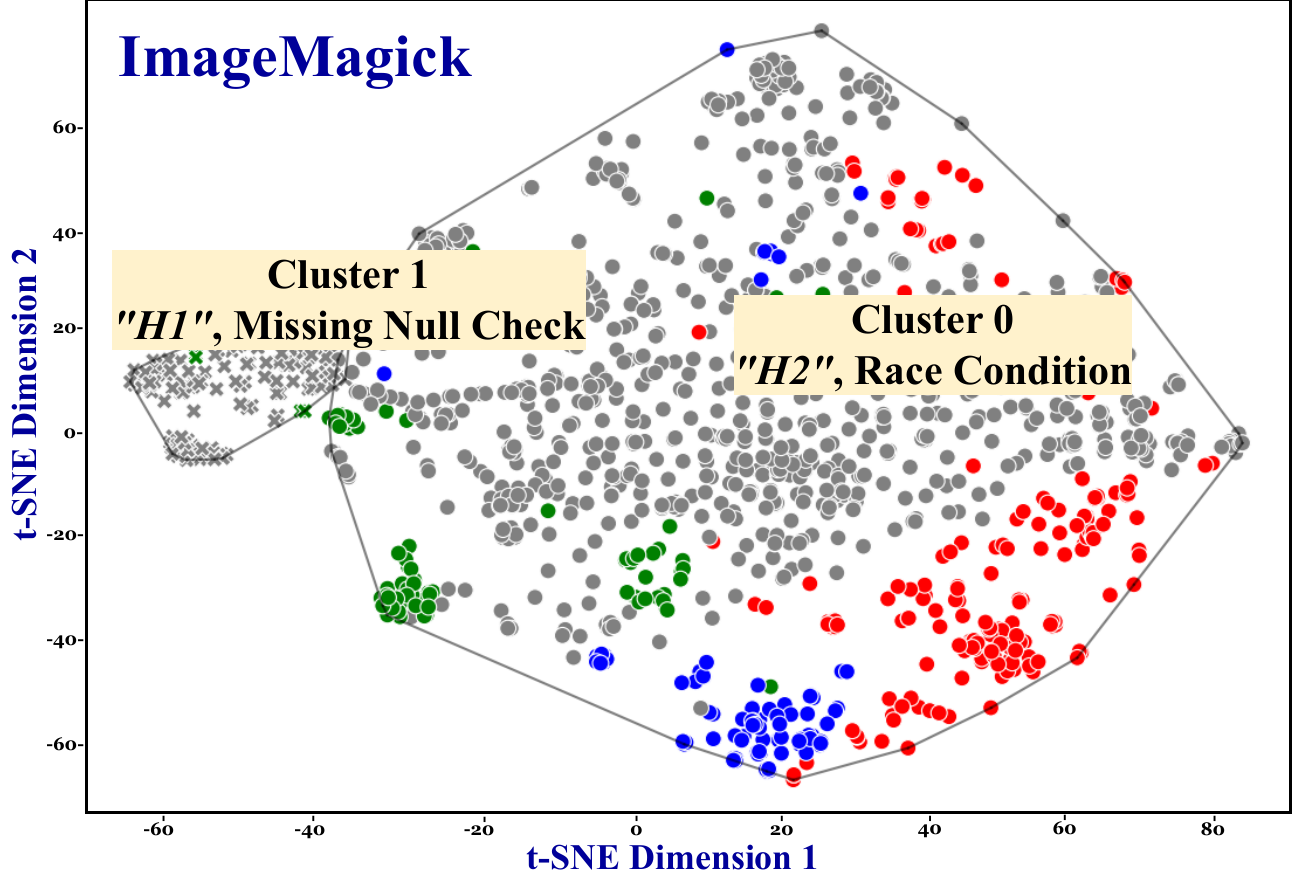}
    \label{fig:imageMagick_t-sne}
  \end{subfigure}
  \par\medskip
  \begin{subfigure}[t]{0.8\textwidth}
    \includegraphics[width=\textwidth, height=0.45cm]{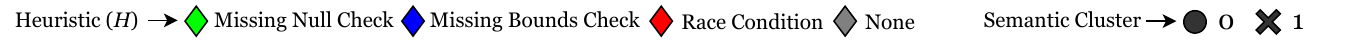}
    \label{fig:legent}
  \end{subfigure}
  \vspace{-0.5cm}
   \caption{\small {Distribution of \texttt{None} labeled patched functions across semantic embedding clusters for each project by HYDRA. Although \texttt{None} cases dominate each cluster, they tend to align with a specific heuristic type H (annotated), indicating latent similarity to known vulnerability patterns within that semantic cluster.}}
   
   \Description{t-SNE visualizations of semantic embeddings for patched functions
    from Chrome, Android, and ImageMagick produced by HYDRA. Each plot shows
    clusters dominated by functions labeled as None, with annotations indicating
    their proximity to specific heuristic vulnerability types, suggesting latent
    similarity to known risky patterns within each semantic cluster.}

   \label{fig:hydra_tsne}
   \vspace{-0.2cm}
\end{figure*}

\vspace{0.5cm}
\subsection{RQ2: Impact of HYDRA’s Components on Clustering Patched Functions}
\textbf{Method.} While RQ1 focuses on model accuracy in predicting known risky patterns, RQ2 shifts focus to understanding how individual components of HYDRA influence the quality of learned internal representations of patched functions used for vulnerability inference. To assess this, we examine how well the function embeddings from each model variant form meaningful clusters in an unsupervised setting. This analysis will help us assess whether HYDRA captures meaningful structure in patched function without relying on explicit vulnerability labels. If distinct clusters emerge, it suggests HYDRA has learned to organize functions based on latent risk patterns even when no known bug type is specified. Together, RQ1 and RQ2 demonstrate HYDRA’s ability to both predict and group subtle post-patch security flaws. To evaluate HYDRA’s overall performance and understand the role of each model component, we compare baseline variants and HYDRA using the unsupervised clustering metrics introduced in Section \ref{sec:metrics}, which assess structural quality (compactness and separation) independently of labeled data. This comparison reveals the incremental contribution of symbolic heuristics, pretrained embeddings, and VAE-based representation learning in shaping the internal structure of the learned space.\\
\textbf{Results:} Table \ref{tbl:score} presents the unsupervised clustering quality metrics such as Silhouette Score ($\uparrow$), CH Index ($\uparrow$), and DB Index ($\downarrow$) for all models across the Chrome, Android, and ImageMagick test projects. 
HYDRA consistently outperforms all other models in unsupervised clustering quality across all projects, achieving the highest Silhouette scores, maximum CHI values and the lowest DBI scores. These results indicate that HYDRA forms well-separated, cohesive clusters in latent space, even without heuristic supervision during inference.
Although M1 (regex + K-Means) achieves perfect clustering scores, this is due to its trivial binary input, offering no real semantic insight or generalization. M2, using only GraphCodeBERT embeddings, performs better on clustering metrics but forms less meaningful clusters. In contrast, M3 (regex + embeddings) captures richer vulnerability structure through symbolic semantic fusion. HYDRA  surpasses all models, with its VAE based encoding producing compact, well-separated, and semantically meaningful clusters enabling the prediction of subtle post-patch risks that simpler models overlook. 
\begin{table}[ht]
\centering
\footnotesize
\caption{Unsupervised clustering quality of models across all projects using Silhouette, CHI and DBI metrics.}
\vspace{-0.2cm}
\label{tbl:score}
\begin{center}

\begin{tabular}{c | c | c | c}
 \hline
  \textbf{Model} & \textbf{Silhouette} \textbf{($\uparrow$)} & \textbf{CHI} \textbf{($\uparrow$)} & \textbf{DBI} \textbf{($\downarrow$)} \\ 
 \hline
 \multicolumn{4}{c}{\textbf{Chrome}} \\
 \hline
 M1 & 1.00 & 1.00 & 0.00 \\ 
 
 M2 & 0.15 & 3405.30 & 2.05 \\
 
 M3 & 0.14 & 3388.33 & 2.06 \\
 
 \shortstack{\textbf{HYDRA}} & \textbf{0.54} & \textbf{21940.15} & \textbf{0.81} \\
 \hline

  \multicolumn{4}{c}{\textbf{Android}} \\
  \hline
  M1 & 0.99 & 1.00 & 0.00 \\ 
 
  M2 & 0.16 & 376.34 & 2.43 \\
 
  M3 & 0.16 & 371.86 & 2.45 \\
 
 \shortstack{\textbf{HYDRA}} & \textbf{0.61} & \textbf{3802.89} & \textbf{0.72} \\
 \hline

  \multicolumn{4}{c}{\textbf{ImageMagick}}\\
  \hline
  M1 & 0.99 & 1.00 & 0.00 \\ 
 
  M2 & 0.16 & 283.36 & 2.37 \\
 
  M3 & 0.15 & 278.94 & 2.40 \\
 
 \shortstack{\textbf{HYDRA}} & \textbf{0.68} & \textbf{1993.26} & \textbf{0.66} \\
 \hline
\end{tabular}
\end{center}
\end{table}

\begin{figure}[ht]
\centering
\includegraphics[width=1 \linewidth]{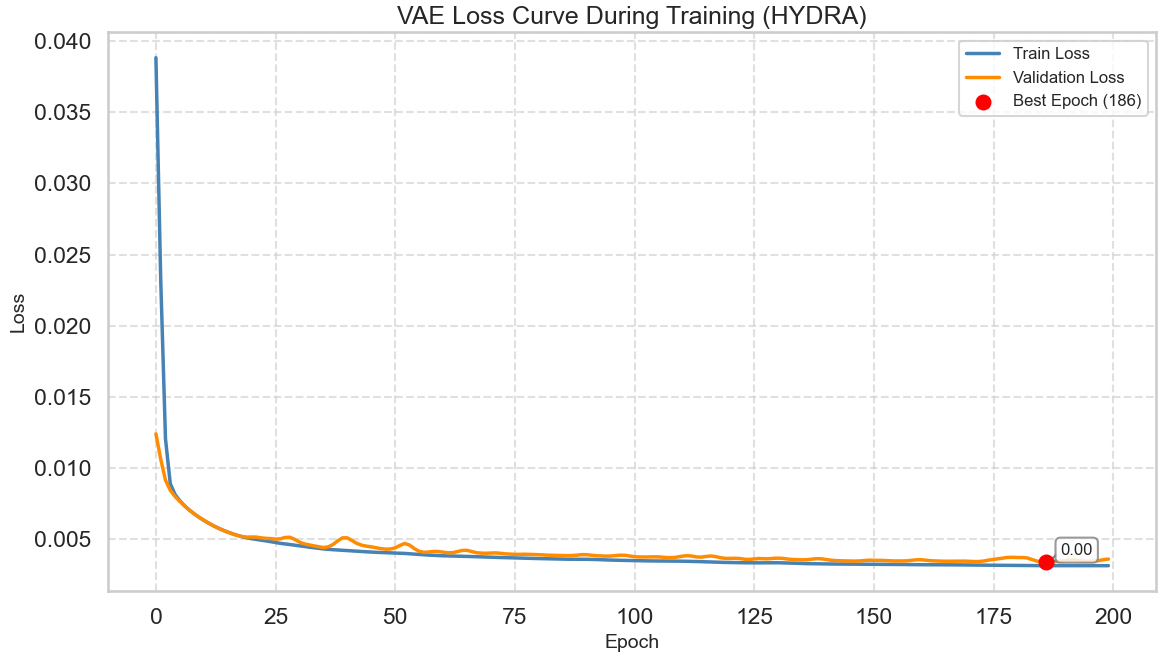}
\caption{VAE reconstruction loss during HYDRA training. Both training and validation losses converge smoothly, with best validation performance at epoch 186.}

\Description{Line plot showing variational autoencoder reconstruction loss during HYDRA training. Training and validation losses decrease smoothly and converge over epochs, with the lowest validation loss occurring around epoch 186.}
\label{fig:vae_loss_curve}
\vspace{-0.5cm}
\end{figure}

\indent Interestingly, all embedding based models (M2, M3, HYDRA) consistently form two semantic clusters across projects, each dominant by \texttt{None} labeled functions.  Figure \ref{fig:hydra_tsne} shows the semantic clustering results derived from HYDRA, which illustrates distribution of \texttt{None} labeled patched functions across semantic embedding clusters for each project. While \texttt{None} cases dominate these clusters, they consistently align with specific heuristic types ($H$), suggesting latent similarities to known vulnerability pattern within each semantic cluster. This suggests that most of the unflagged functions often share latent structural traits like control flow or data usage and this behavior is likely influenced by the GraphCodeBERT embeddings which encode rich semantic and syntactic relationship. HYDRA, in particular, surfaces these high risk, anomalous patterns that may escape rule based prediction and pose zero-day risks. In contrast, M1 fails to form such clusters, underscoring the limitations of purely symbolic methods and the representational strength of HYDRA’s hybrid approach.
\vspace{0.05cm}
\noindent \textbf{Effectiveness:} HYDRA outperforms all variants across clustering metrics by combining Regex rules, GraphCodeBERT embeddings, and VAE-based latent learning, yielding more coherent and semantically meaningful clusters.To further analyze this improvement, we examine HYDRA's VAE reconstruction loss. As depicted in Figure \ref{fig:vae_loss_curve}, the model converges rapidly within the first few epochs and maintains stable error reduction across 200 epochs. The lowest validation loss occurs at epoch 186, approaching zero. This stability confirms that HYDRA’s VAE effectively learns a low dimensional latent space that retains both heuristic and semantic structure. This latent alignment, combined with downstream K-means clustering, enables HYDRA to deliver strong prediction performance and structural generalization, as evidenced in RQ1 and RQ2.

\vspace{-0.2cm}
\subsection{RQ3: HYDRA’s Capacity to Identify Novel/Unseen Risk Patterns}
\textbf{Method.}  
To evaluate HYDRA’s ability to generalize beyond predefined heuristic rules, we focus on its predictions labeled as \texttt{None} cases not matching any of the five heuristics. A high proportion of such cases in every cluster across all the projects as seen in Figure \ref{fig:hydra_tsne} suggests  HYDRA's potential to uncover latent risks.
Building on RQ2, which showed that HYDRA forms well-separated, meaningful clusters, RQ3 examines how these \texttt{None} cases  tend to align with a specific heuristic type H of known vulnerabilities. This suggests HYDRA can uncover overlooked or emerging vulnerability patterns that may signal future zero-day threats.
\begin{table}[ht]
\centering
\footnotesize
\caption{Prediction of \texttt{None} labeled samples from all three model variants. Values indicate total counts and corresponding percentages.}
\vspace{-0.2 cm}
\label{tbl:none_type}
\begin{tabular}{c | c | c | c}
 \hline
 \textbf{Model} & Chrome & Android & ImageMagick\\ 
 \hline
 M1 & 0 & 0 & 0\\
 M3 & 15446 (94.3\%) & 2018 (86.9\%) & 1352 (79.4\%)\\
 \textbf{HYDRA} & \textbf{15599} (95.2\%) & \textbf{2065} (88.9\%)& \textbf{1430} (83.97\%)\\
 \hline
\end{tabular}
\end{table}
\\
\noindent \textbf{Results:} 
Table \ref{tbl:none_type} reports the proportion of samples labeled as \texttt{None} across the three model variants. As expected, the regex only model (M1) produces no \texttt{None} predictions due to its deterministic design it can only match explicitly defined patterns, limiting its ability to surface novel or ambiguous risks, hence 0 count. In contrast, the latent space models M3 and HYDRA predict substantial proportions of \texttt{None} labels, with HYDRA reaching 95.2\% in Chrome, 88.9\% in Android, and 83.97\% in ImageMagick. This reflects HYDRA’s greater generalization capacity beyond handcrafted rules.

To assess the plausibility of these \texttt{None} predictions, we analyze their distribution in HYDRA's latent space and identify the known heuristic it aligns the most with in each of the two distinct semantic clusters 0 and 1. Table \ref{tab:probable_region} highlights the most probabilistically aligned/dominant heuristic type $H_{A}$ in each of the clusters for \texttt{None} labeled samples across Chrome, Android, and ImageMagick based on the highest softmax confidence scores produced by HYDRA’s heuristic prediction head. For example, in the Chrome project, most \texttt{None} cases align with two dominant heuristics: 934 samples from cluster 0 map to $H_1$ (CWE-476: NULL pointer dereference), and 530 samples from cluster 1 resemble $H_2$ (CWE-362: Thread non-synchronization). Similar dual-pattern alignments emerge in Android ($H_2$, $H_3$) and ImageMagick ($H_1$, $H_2$). These probabilistic associations arise from HYDRA’s latent space. Furthermore we manually inspected a representative subset of \texttt{None} predicted functions with two independent software engineers (K = 0.89), observing that HYDRA’s latent space probabilistically clusters structurally and semantically similar functions near regions associated with known heuristics, even without explicit matches. Rather than hard relabeling, this behavior offers a soft signal of latent risk, surfacing functions potentially indicative of emerging or uncharacterized vulnerabilities. These associations show HYDRA’s potential to surface risky code patterns missed by symbolic methods, aiding zero-day vulnerability prediction. By aligning with heuristic and CWE contexts, HYDRA enhances interpretability and supports informed triage even without explicit rule matches.\\

\vspace{-0.2cm}

\vspace{-0.3cm}
\begin{table}[ht]
\centering
\footnotesize
\caption{Most dominant heuristic aligned types ($H_A$) for \texttt{None} labeled samples in Cluster 0 and Cluster 1 across Chrome, Android, and ImageMagick, as positioned in HYDRA’s latent space.}
\begin{tabular}{m{1.25cm} | m{1.2cm} | m{0.6cm} | m{1cm} | m{2.8cm}}
\hline
\centering \textbf{Project} & \centering \textbf{Cluster({$H_A$})} & \centering \centering \textbf{\texttt{None} Count} & \centering \textbf{CWE Match} & \centering \arraybackslash 
 \textbf{Interpretation} \\
\hline
\centering \multirow{2}{*}{Chrome} & \centering 0 ($H_1$) & \centering 934 & \centering CWE-476 & NULL dereference \\
                        & \centering 1 ($H_2$) & \centering 530 & \centering CWE-362 & Thread non-synchronization \\
\hline
\centering \multirow{2}{*}{Android} & \centering 0 ($H_2$) & \centering 157 & \centering CWE-362 & Thread non-synchronization \\
                         & \centering 1 ($H_3$) & \centering 65  & \centering CWE-119 & Buffer overflow \\
\hline
\centering \multirow{2}{*}{ImageMagick} & \centering 0 ($H_1$) & \centering 29  & \centering CWE-476 & NULL dereference \\
                             & \centering 1 ($H_2$) & \centering 106 & \centering CWE-362 & Thread non-synchronization \\
\hline
\end{tabular}
\label{tab:probable_region}
\end{table}

\vspace{-0.5cm}
\section{DISCUSSION}
\label{sec:dis}
\begin{figure}[htbp]
  \centering
  \setlength{\fboxsep}{1pt}
  \setlength{\fboxrule}{0.5pt}
  \fcolorbox{gray!60}{white}{
  \includegraphics[width=7cm, height=1.5cm]{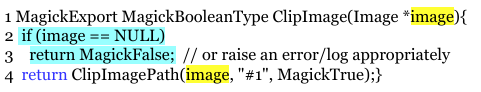}
  }
  \caption{Example from ImageMagick: HYDRA flags the \texttt{ClipImage} wrapper function using heuristic H1 (missing null check).}

  \Description{Code snippet from ImageMagick showing the ClipImage wrapper function flagged by HYDRA due to a missing null check heuristic.}
  \label{fig:discussion_example1}
  \vspace{-0.5cm}
\end{figure}

To further validate HYDRA’s practical impact in the context of software security testing, we analyzed patched functions by identifying risky patterns through two real-world case studies from the Chrome and ImageMagick software projects.\newline 
\textbf{ImageMagick}. A representative case from the ImageMagick project (Figure \ref{fig:discussion_example1}) shows that, despite patching, subtle issues like unchecked null pointer usage can persist posing latent risks for future zero-day vulnerabilities. In this case, the wrapper function \texttt{ClipImage} uses the input pointer \texttt{image} (marked yellow, line 1 and 4) without verifying its validity, risking a null dereference. HYDRA identifies such post-patch flaws by combining symbolic heuristics with learned code representations, revealing incomplete fixes often missed by manual review or traditional scanners.\\
\indent \textbf{Possible Corrected Version:} A safeguard is added (marked cyan, lines 2–3 from Figure \ref{fig:discussion_example1}) to ensure the pointer is not null before use, effectively mitigating the vulnerability. Such missing null checks can result in undefined behavior, application crashes, or denial-of-service conditions when dereferenced improperly. This issue corresponds to CWE-476 (NULL Pointer Dereference) and poses significant risks in systems processing untrusted inputs, potentially leading to memory corruption or enabling exploit chains.
\vspace{-0.5cm}

\begin{figure}[htbp]
  \centering
  \setlength{\fboxsep}{1pt}
  \setlength{\fboxrule}{0.5pt}
  \fcolorbox{gray!60}{white}{
  \includegraphics[width=7cm, height=3.6cm]{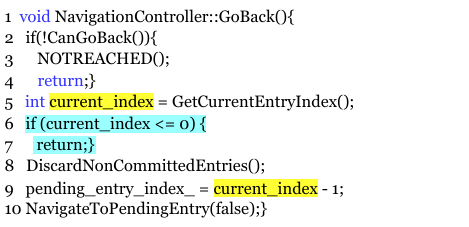}
  }
  \caption{Example from Chrome: \texttt{GoBack} function is marked as \texttt{None} by HYDRA but semantically aligns with heuristic H3 (missing bounds check).}
  
  \vspace{-0.3cm}
  \Description{Code example from Chrome showing the GoBack function labeled as None by HYDRA but exhibiting behavior similar to a missing bounds check heuristic.}
  \label{fig:discussion_example2}
  \vspace{-0.3cm}
\end{figure}

\noindent \textbf{Chrome.} Another example from the Chrome project (see Figure \ref{fig:discussion_example2}) shows a function labeled as \texttt{None} by HYDRA. Although the expression: \texttt{pending\_entry\_index\_=current\_index-1} suggests a potential underflow risk, HYDRA's $H_3$ heuristic did not trigger. This is because $H_3$ targets recognizable patterns of unsafe index use, such as: (1) unvalidated index access (e.g.\texttt{array[i]}, \texttt{vector.at(i)}, \texttt{buffer[i])}, and (2) index comparisons (e.g., $i < size$, $i >= 0$). Since this case involves a simple assignment without dereference or container access, the symbolic rule was not matched. Though \texttt{GoBack()} did not match HYDRA’s symbolic $H_3$ heuristic, its latent embedding placed it near $H_3$ labeled samples due to structural similarities specifically, decrementing an index without bounds checking and using it in a downstream state transition (marked yellow, line 5 and 9).\\
\indent \textbf{Possible Corrected Version:} There has been added a bounds check (marked cyan, lines 6–7, in Figure \ref{fig:discussion_example2}) such as \texttt{if(current\_ind-\newline ex <= 0)}, preventing underflow and mitigating this latent vulnerability. HYDRA’s learned representation correctly associates this with $H_3$ style risks, highlighting its ability to flag latent issues beyond explicit rule matches.
This unsafe arithmetic aligns with CWE-1285: Improper Validation of Specified Index, where missing lower bound checks can lead to invalid states or out-of-bounds behavior. Such subtle flaws often evade conventional testing and, under certain conditions, may be exploitable as zero-day vulnerabilities.

\indent These case studies highlight HYDRA’s ability to predict latent both known and unknown vulnerabilities by combining symbolic rules with deep code representations. This fusion enables HYDRA to surface subtle post-patch flaws that symbolic methods miss, making it a valuable tool for a robust software security testing leveraging zero-day risk prediction.

\vspace{-0.2cm}
\section{THREATS TO VALIDITY}
\label{sec:val}
\textbf{Internal Validity.} 
A key consideration for internal validity is the design of the heuristic labeling process. HYDRA is trained using five manually defined rules derived from regular expressions over preprocessed source code. While these rules are based on widely accepted and frequently observed vulnerability patterns (e.g., missing null or bounds checks) in real-world CVEs, manual construction may introduce inconsistencies or overlook edge cases, potentially introducing noise into the training set. Importantly, HYDRA does not label \texttt{None} classified functions as definitively safe or vulnerable. Instead, it learns latent representations that capture semantic and structural similarity. When \texttt{None} samples appear near known risky clusters, we interpret this as a soft signal of latent risk, informed by qualitative review not as hard relabeling. This ability to surface structurally similar but heuristically silent code offers practical utility in vulnerability triage and zero-day threat anticipation. Due to the lack of ground \textbf{truth labels} for post-patch vulnerability assessment, we did not perform direct comparative analysis and compute conventional metrics (e.g., F1, AUC, MCC), focusing instead on unsupervised vulnerability prediction and heuristic alignment.
\\
\textbf{External Validity:} 
Our dataset consists of patched (non-vulnerable) C/C++ functions sourced from Linux-based repositories in the Big-Vul dataset. While C/C++ are widely studied in software security, HYDRA has not yet been evaluated on other languages like Java, Python, or Rust, limiting generalizability across different programming paradigms. HYDRA produces both heuristic class predictions and a \texttt{None} class, indicating that a function does not match any predefined rule potentially signaling a zero-day or novel pattern. However, without a definitive ground truth for \texttt{None} cases, this interpretation remains probabilistic. While it aligns with the goal of uncovering latent vulnerabilities, quantitative evaluation of \texttt{None} predictions is challenging. These cases were partially validated through expert review and manual inspection. Future work will explore stronger validation signals, such as fuzzing based analysis or synthetic vulnerability injection.

\section{RELATED WORK}
\label{related_work}
\textbf{Vulnerability Detection in Source Code.} Extensive research has focused on deep learning models trained on large vulnerability datasets to detect insecure code. Approaches like VulDeePecker \cite{VuldeePecker}, DeepCVA \cite{DeepCVA}, and LineVul \cite{LineVul} use token-level or structured embeddings to classify code as vulnerable or not. More recent methods such as hybrid models like IVDetect \cite{IVDetect} and ReVeal \cite{Reveal}, and graph based models like Devign \cite{Devign} leverage commit-level changes and control/data flow representations to improve generalization. However, most of these rely on explicitly labeled buggy code, limiting their ability to detect latent or residual risks in previously patched function an area that HYDRA is specifically designed to address.\\
\textbf{Heuristic and Static Rule Based Detection.} Conventional tools like Cppcheck \cite{Cppcheck} and Flawfinder \cite{Flawfinder} rely on manually crafted rules or regex to detect issues such as use-after-free, buffer overflows, and missing null checks. While interpretable, these tools often lack precision and fail to generalize to subtle, latent vulnerabilities in real-world code. In contrast, HYDRA integrates such heuristics as structured feature vectors, guiding deep representations toward known vulnerability patterns.\\
\textbf{Patch Analysis and Silent Fix Studies.} Recent efforts like GraphSPD \cite{GraphSPD}, PatchRNN \cite{PatchRNN}, and GRAPE \cite{Grape} have begun analyzing post-fix code to detect silent fixes and assess patch security. These methods often rely on change classification or learning pre- and post-patch code representations. However, they typically assume the patch is correct or focus on binary classification of patch types. In contrast, HYDRA evaluates previously patched function to uncover residual risks, identifying cases where fixes are incomplete or incorrectly applied.\\
\textbf{Hybrid and Casual Models in Vulnerability Learning.} Hybrid systems like DeepDFA \cite{DeepDFA} and CausalVul \cite{CasualVul} combine symbolic reasoning with deep learning to improve generalization beyond spurious correlations. HYDRA builds on this direction by integrating human interpretable heuristics with deep semantic representations from GraphCodeBERT, forming a unified architecture that balances precision and explainability for detecting silent vulnerabilities in post-fix code.\\
\textbf{Zero-Day and Out-of-Distribution (OOD) Prediction.} To combat zero-day threats, approaches like NERO \cite{nero}, UL-VAE \cite{ul-vae}, and open-set intrusion detection systems use anomaly or out-of-distri-\newline bution (OOD) detection, typically over binary data or traffic logs via autoencoders or meta-learning. In contrast, HYDRA is tailored for source code and requires neither CVE labels nor runtime triggers enabling probabilistic forecasting of latent zero-day risks directly from patched functions.

\section{CONCLUSION \& FUTURE WORK}
\label{sec:conc}
In this work, we introduce HYDRA, a hybrid vulnerability analysis framework designed to identify residual risky patterns in patched functions potential indicators of zero-day vulnerabilities. HYDRA leverages a dual layered approach: combining handcrafted heuristic rules (regex-based) with deep code representations learned via GraphCodeBERT and a Variational Autoencoder. This integration allows the system to surface latent, high risk patterns that persist even after developers issue security patches patterns that might otherwise evade detection using traditional static or learned techniques in isolation. Our empirical evaluation across three real-world projects Chrome, Android, and ImageMagick shows that HYDRA reduces overmatching compared to heuristic only baselines while retaining the ability to flag vulnerable code segments that may still harbor security weaknesses. The cross project generalization capabilities highlight HYDRA's effectiveness in spotting common coding flaws that survive across diverse codebases. These findings underscore the pressing need to revisit post patch code with a deeper, hybrid analysis lens, as even patched functions can remain susceptible to exploitation a characteristic often seen in zero-day scenarios. As future work, we plan to expand HYDRA by incorporating a richer set of heuristics and evaluating on diverse datasets across languages and domains, including android and cloud systems, to better assess post-patch zero-day risks while further integrating it into automated security testing pipelines to improve post-patch zero-day vulnerability assessment and validation.

\bibliographystyle{ACM-Reference-Format}
\bibliography{sample-sigconf}


\end{document}